\documentclass[onecolumn,showpacs,nobibnotes,nofootinbib]{revtex4}

\usepackage{amsmath,amssymb}
\usepackage{graphicx}
\usepackage{subfigure}

\newcommand{\N}{{\cal{N}}}
\newcommand{\tN}{\tilde {\cal N}}
\newcommand{\x}{\mathbf{x}}
\newcommand{\y}{\mathbf{y}}
\newcommand{\z}{\mathbf{z}}
\newcommand{\vecr}{\mathbf{r}}
\newcommand{\vecb}{\mathbf{b}}
\newcommand{\vrho}{\mathbf{r}}
\newcommand{\vk}{\mathbf{k}}
\newcommand{\vq}{\mathbf{q}}
\newcommand{\vp}{\mathbf{p}}
\newcommand{\be}{\begin{equation}}
\newcommand{\ee}{\end{equation}}
\newcommand{\bea}{\begin{eqnarray}}
\newcommand{\eea}{\end{eqnarray}}
\newcommand{\f}{\frac}

\newcommand{\abs}[1]{\left|#1\right|}

\begin{document}

\title{The Balitsky-Kovchegov equation in full momentum space}
\author{C. Marquet}
\email{marquet@spht.saclay.cea.fr}
\author{G. Soyez\footnote{on leave from the fundamental theoretical physics 
group of the University of Li\`ege.}}
\email{gsoyez@spht.saclay.cea.fr}
\affiliation{SPhT \footnote{URA 2306, unit\'e de recherche associ\'ee au CNRS.}, 
CEA Saclay, B\^{a}t. 774, Orme des Merisiers, F-91191 Gif-Sur-Yvette, France}

\begin{abstract}
We analyse the Balitsky-Kovchegov (BK) saturation equation in momentum space 
and solve it numerically. We confirm that, in the limit 
where the transverse momentum of the incident particle $k$ is much bigger than 
the momentum transfer $q,$ the equation admits traveling-wave solutions. We 
extract the $q$ dependence of the saturation scale $Q_s(Y)$ and verify that 
$Q_s(Y\!=\!cste)$ scales as $\max(q,Q_T),$ where $Q_T$ is the scale caracterizing the
target. 
\end{abstract}

\maketitle

\section{Introduction}\label{sec:intro}

The problem of understanding scattering near the unitarity limit in perturbative
QCD has received a lot of attention over the past ten years. In the Regge limit 
of perturbative QCD, {\em i.e.} when the centre-of-mass energy in a collision is 
much bigger than the fixed hard scale of the problem, parton densities inside the 
projectiles grow with increasing energy, leading to the growth of the scattering 
amplitudes. As long as the densities are not to high, this growth is described 
by the BFKL equation~\cite{bfkl}. But as the parton density becomes higher and 
the scattering amplitude approaches the unitarity limit, one enters in a regime 
called saturation~\cite{glr, glr+}, where BFKL evolution breaks down.

To study scattering amplitudes in the Regge limit, the $q\bar q$-dipole model 
\cite{dipole,mueller} has been developed. This formalism constructs the 
light-cone wavefunction of a dipole (a quark-antiquark pair in the colour singlet 
state) in the leading logarithmic approximation. The size of the dipole provides 
the hard scale that justifies the use of perturbation theory. As the energy 
increases, the dipole evolves and the wavefunction of this evolved dipole is 
described as a system of elementary dipoles. When this system of dipoles 
scatters on a target, the scattering amplitude has been shown to obey the BFKL 
equation. 

Interestingly enough, the dipole formalism was shown to be also well-suited to 
include density effects and non-linearities that lead to saturation and 
unitarization of the scattering amplitudes. Indeed, Kovchegov~\cite{kov} derived 
an equation for the dipole scattering amplitude that reduces to the BFKL 
equation when the amplitude is small and that contains high-energy unitarity 
effects as the amplitude reaches unity. Although this equation is 
not exact, as it is a mean field approximation of the more complete infinite 
hierarchy of equations established by Balitsky~\cite{bal}, it has the advantage 
of being a closed equation for the scattering amplitude. This equation, which is 
usually referred to as the BK (Balitsky-Kovchegov) equation, is propably
the simplest and most accurate equation one can consider to describe the saturation
regime of QCD.

The BK equation is an evolution equation for $\N(\vecr,\vecb,Y),$ the imaginary 
part of the scattering amplitude of a dipole of transverse size $\vecr$ at 
impact parameter $\vecb$ and where the rapidity $Y$ is the logarithm of the 
centre-of-mass energy. It is a non-linear equation that resums QCD fan diagrams 
in the leading-logarithmic approximation~\cite{kov}. Many interesting features 
of this equation have been understood analytically~\cite{bk+,Munier1} for the $\vecr$ 
and $Y$ dependences. In particular, it was shown that the asymptotic 
solutions have the so-called geometric scaling \cite{geomsc} property 
$\N(\vecr,\vecb\!\sim\!0,Y)\!=\!\N(|\vecr|Q_T\Omega_s(Y)),$ where $Q_T$ is a 
scale characterising the target and $\Omega_s(Y)\!=\!\exp(vY)$ is the dimensionless
saturation scale. One has not been able however to extract any useful information on  
the $\vecb$ dependence. The analysis of the BK equation in impact-parameter space
(see {\em e.g.} ~\cite{conf,bknum}) even shows a contradiction with confinement as the 
large-$\vecb$ dependence of the solutions develops a power-law tail with increasing 
rapidity.

It was recently proposed~\cite{us} to analyse the equation in momentum space, by 
Fourier transforming $\N(\vecr,\vecb,Y),$ into $\tilde{\N}(\vk,\vq,Y)$ where 
$\vk$ can be interpreted as the incoming transverse momentum of the dipole and 
$\vq$ as the momentum transfer. It was shown in~\cite{us} that the three 
variables $\vk,$ $\vq,$ and $Y$ give information on the approach to the unitarity 
limit. In particular, it was predicted that the geometric scaling property should 
also be present in momentum space when $|\vk|\!\gg\!|\vq|$ as one should have 
asymptotically $\tilde{\N}(\vk,\vq,Y)\!=\!\tilde{\N}[|\vk|/(|\vq|\Omega_s(Y))].$ 
The goal of this paper is to formulate and analyse the BK equation in momentum space, 
to solve it numerically and compare our results with those predicted. The equation in
momentum space has already been studied numerically in the $\vecb$-independent situation 
\cite{bknumimp}. Our purpose here is to extend those analysis to full 
momentum space including the transfer momentum dependence.

The plan of the paper is as follows. In section \ref{sec:bdep}, we write the BK 
equation in momentum space and derive some analytical properties. 
In section \ref{sec:trav}, we briefly explain the properties of asymptotic solutions 
and how they translate into geometric scaling. In Section \ref{sec:numsol}, we give 
details about the numerical techniques used to solve the equation and present our 
results. Section \ref{sec:ccl} concludes.

\section{The BK equation in momentum space}\label{sec:bdep}

\begin{figure}
  \subfigure[]{\includegraphics{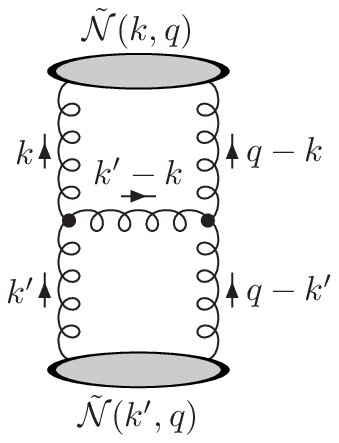}}\hspace{2.0cm}
  \subfigure[]{\includegraphics{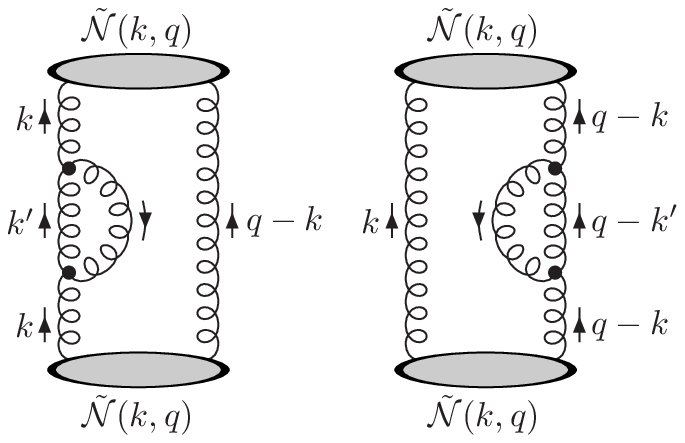}}
\caption{Linear contributions to the evolution of the dipole density in momentum 
space: (a) real gluon emission, (b) virtual gluon emission.}\label{fig:linterms}
\end{figure}

\begin{figure}
\includegraphics{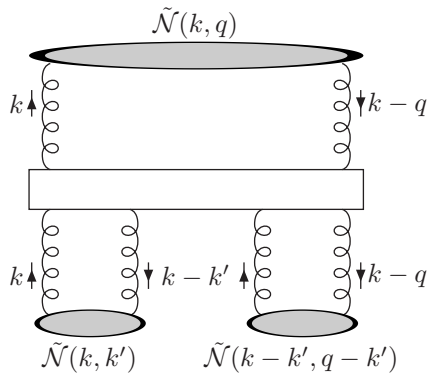}
\caption{Non-linear contribution to the evolution of the dipole density in 
momentum space. This correspond to the resummation of fan 
diagrams.}\label{fig:nonlin}
\end{figure}

The BK equation in coordinate space reads
\be\label{eq:bkspace}
\partial_Y \N(\x,\y) = \frac{\bar{\alpha}}{2\pi}\int d^2z 
\frac{(\x\!-\!\y)^2}{(\x\!-\!\z)^2(\z\!-\!\y)^2} 
\left[\N(\x,\z)+\N(\z,\y)-\N(\x,\y)-\N(\x,\z)\N(\z,\y) \right]
\ee
where $\bar\alpha\!=\!\alpha_sN_c/\pi.$ The two-dimensional vectors $\x$ and 
$\y$ are the transverse positions of the quark and antiquark composing  the 
dipole of transverse size $\vecr\!=\!\x\!-\!\y$ at impact parameter 
$\vecb\!=\!(\x\!+\!\y)/2.$ The rapidity $Y$ is not mentioned explicitly in the 
amplitude but $\N(\x,\y)$ always carries a $Y$ dependence. Let us define the 
following Fourier transformation:
\begin{equation}\label{eq:fourier}
\tilde{\N}(\vk,\vq)=\f1{(2\pi)^2}\int d^2x\,d^2y\: e^{i\vk.\x}e^{i(\vq-\vk).\y}
\,\f{\N(\x,\y)}{(\x\!-\!\y)^2}\ .
\end{equation}
Then the equation verified by $\tilde{\N}(\vk,\vq)$ can be written as
\be\label{eq:fullbk}
\partial_Y \tN(\vk,\vq) 
 = \frac{\bar{\alpha}}{\pi} \int \f{d^2k'}{(\vk\!-\!\vk')^2} 
 \left\{ \tN(\vk',\vq)-\f1{4} \left[
   \f{(\vq\!-\!\vk)^2}{(\vq\!-\!\vk')^2}+
   \f{\vk^2}{\vk'^2} 
   \right]\tN(\vk,\vq)
 \right\}
 -\frac{\bar{\alpha}}{2\pi}\int d^2k'\ 
\tN(\vk,\vk')\tN(\vk\!-\!\vk',\vq\!-\!\vk').
\ee
This equation involves three contributions. The first integration contains the 
linear terms corresponding to real and virtual gluon emissions (see 
Fig.\ref{fig:linterms}). The last term is the nonlinear correction which, as 
represented in Fig.\ref{fig:nonlin}, arises from the resummation of fan diagrams. 
A trivial change of variable in the virtual correction term in Eq. 
\eqref{eq:fullbk} can put the evolution equation under a form emphasizing more 
clearly the cancellation of divergences between the real and virtual gluon 
emissions. We get
\begin{eqnarray}
\partial_Y \tN(\vk,\vq)
 & = &\frac{\bar\alpha}{\pi}\int \frac{d^2k'}{(\vk-\vk')^2}
   \left\{
   \tN(\vk',\vq) - \frac{1}{2} \left[
     \frac{\vk^2}{\vk'^2+(\vk-\vk')^2} +
     \frac{(\vq-\vk)^2}{(\vq-\vk')^2+(\vk-\vk')^2}
     \right] \tN(\vk, \vq)
   \right\} \nonumber \\
 & - & \frac{\bar\alpha}{2\pi} \int 
d^2k'\,\tN(\vk,\vk')\,\tN(\vk-\vk',\vq-\vk').
\label{eq:bk}
\end{eqnarray}

Let us now look at the properties of this equation in some particular cases. 
If the amplitude (in coordinate space) only depends on the size of the dipole 
{\em i.e.} $\N(\x,\y)\!=\!\N(\x\!-\!\y)$, then 
$\tN(\vk,\vq)\!=\!2\pi\delta^{(2)}(\vq)\tN(\vk)$, where
\be
\tN(\vk)=\int\f{d^2x}{2\pi\x^2}\ e^{i\vk.\x}\N(\x)
\ee
obeys
\begin{equation}\label{eq:bkbindep}
\partial_Y \tN(\vk) = \f{\bar{\alpha}}{\pi} \int \f{d^2k'}{(\vk\!-\!\vk')^2}
\left[ \tN(\vk')-\f{\vk^2}{\vk'^2+(\vk-\vk')^2}\tN(\vk) \right]
-\bar{\alpha}\,\tN^2(\vk).
\end{equation}
One recovers the usual equation \cite{glr} discussed for $\vecb$-independent 
situations.\\
Let us now look at the equation in the forward limit $\vq\to 0$. Due 
to the fact that the non-linear term in \eqref{eq:bk} is off-diagonal, the 
equation for the forward dipole amplitude involves non-forward contributions. 
However, if we assume that the amplitude is peaked around $|\vq|=0$ and has the 
factorised form
\begin{equation}\label{eq:facto}
\tN(\vk, \vq) = \phi(|\vq|)\,\tN_0(\vk),
\end{equation}
the non-linear term can be simplified
\[
\int d^2k'\:\tN(\vk,\vk')\,\tN(\vk-\vk',\vq-\vk') \approx 2\pi\,\tN_0^2(\vk) 
\int_0^\infty k'\,dk'\:\phi^2(k'). 
\]
If we normalize the $\vq$-dependent factor in \eqref{eq:facto} in such a way that
\[
\phi(0) = \int_0^\infty k'\,dk'\:\phi^2(k'),
\]
we obtain a closed equation for the rapidity evolution of $\tN_0(\vk)$ of the form
\[
\partial_Y \tN_0(\vk) = \f{\bar{\alpha}}{\pi} \int \f{d^2k'}{(\vk\!-\!\vk')^2}
\left[ \tN_0(\vk')-\f{\vk^2}{\vk'^2+(\vk-\vk')^2}\tN_0(\vk) \right]
-\bar{\alpha}\,\tN_0^2(\vk),
\]
which is equivalent to the equation \eqref{eq:bkbindep} obtained in the 
$\vecb$-independent situation. It is however important to realize that, while the 
simplification is exact for $\vecb$-independent amplitude, it requires assumptions 
on the small-$\vq$ dependence of the amplitude in the forward limit.

\section{Traveling waves and geometric scaling}\label{sec:trav}

In this section, we briefly summarize the properties of the asymptotic solutions of the BK equation and replace this in the context of the new form \eqref{eq:fullbk}.

Let us start with the case of the $\vecb$-independent equation \eqref{eq:bkbindep}. It has been shown~\cite{Munier1} that, in the diffusive approximation, the equation for $\tN(k\!=\!|\vk|)$ is equivalent to the Fisher-Kolmogorov-Petrovsky-Piscounov (F-KPP) equation~\cite{KPP}. This equation, well known in statistical physics, admits traveling-wave solutions at large rapidities and momentum $k$, which means that the solution is a function of $Y\!-v_c\!\log(k)$ only, where $v_c$ is the speed of the wavefront.
More generally, if the evolution equation satisfies the following conditions:
\begin{itemize}
\item $\tN=0$ is an unstable fix point and the equation contains a linear term, leading to a growth of the amplitude, and a non-linear damping term; this is realized in \eqref{eq:bkspace} 
and \eqref{eq:fullbk},
\item the initial condition is steep enough, which in the QCD case is ensured by colour transparency,
\item the linearised equation admits a superposition of waves as solution 
\begin{equation}\label{eq:waves}
\tN_{\text{lin}}(k,Y) = \int_{c-i\infty}^{c+i\infty} \frac{d\gamma}{2i\pi} a_0(\gamma) \exp\left[\omega(\gamma) Y - \gamma \log(k^2/k_0^2)\right],
\end{equation}
\end{itemize}
then the asymptotic solution of the full equation takes the form
\begin{equation}\label{eq:travwaves}
\tN(k,Y) \underset{Y\to\infty}{\sim} \left[\frac{k^2}{Q_s^2(Y)}\right]^{-\gamma_c},
\end{equation}
for $k\gg k_0$ where the saturation scale $Q_s^2(Y)$ grows as $k_0\ e^{v_c Y}$. The critical exponent $\gamma_c$ and the critical speed $v_c$ are given by
\[
v_c = \frac{\omega(\gamma)}{\gamma} = \omega'(\gamma),
\]
meaning that the group velocity is equal to the phase velocity.

In the case of the BK equation, the corresponding linear equation is the BFKL equation, whose solutions have been extensively studied~\cite{Lipatov86,NP,Navelet:1997tx,LipatovReview}. For the impact-parameter-independent case, Munier and Peschanski have proven~\cite{Munier1} that the three conditions for traveling waves are fulfilled. In a recent study, we have shown~\cite{us} that this can be extended to the full BK equation including momentum transfer. In this case, the reference scale $k_0$ becomes proportional to the typical scale of the target $Q_T$ or to the momentum transfer $|\vq|$ whether $|\vq|$ is smaller or larger than $Q_T$. Before going into the numerical studies, we recall the important points concerning geometric scaling at non-zero momentum transfer. 

In the previous work~\cite{us}, one of the key points was to show that the BFKL solutions can be put under the form \eqref{eq:travwaves} when $|\vk|\!\gg\!|\vq|$. Since the definition \eqref{eq:fourier} of the Fourier-transformed amplitude introduces an additional $(\x\!-\!\y)^2$ w.r.t. the usual one, the solution is slightly different to the one we used in~\cite{us}. However our conclusions are unchanged as we shall see now. Let us start with the well-known result~\cite{Lipatov86,NP,Navelet:1997tx,LipatovReview} that 
\[
\N_{\text{lin}}(\x,\y)=
\int_{\frac{1}{2}-i\infty}^{\frac{1}{2}+i\infty} 
\frac{d\gamma}{2i\pi}\ e^{\bar\alpha\chi(\gamma)Y}
\left(\frac{\abs{\x-\y}}{\abs{\x}\abs{\y}}\right)^{2\gamma}\phi_0(\gamma)
\]
is solution of the BFKL equation in coordinate space with
\begin{equation}\label{eq:bfklchi}
\chi(\gamma) = 2\psi(1)-\psi(\gamma)-\psi(1-\gamma).
\end{equation}
and where $\phi_0(\gamma)$ is a function that specifies the initial condition. Note that we kept only the dominant component at high energy which corresponds to vanishing conformal spin $n=0.$
If we now Fourier transform this solution according to \eqref{eq:fourier}, one obtains
\[
\tN_{\text{lin}}(\vk,\vq)=\int_{\frac{1}{2}-i\infty}^{\frac{1}{2}+i\infty} \frac{d\gamma}{2i\pi}\ e^{\bar\alpha\chi(\gamma)Y}
f^\gamma(\vk, \vq)\phi_0(\gamma,\vq)
\]
with
\begin{equation}\label{eq:linvpr}
f^\gamma(\vk, \vq) = \frac{\Gamma^2(\gamma)}{\Gamma^2\left(\frac{1}{2}+\gamma\right)} \frac{2}{\abs{k}} \left|\frac{q}{4k}\right|^{2\gamma-1}{}_2{\rm F}_1\left(\gamma,\gamma;2\gamma;\frac{q}{k}\right){}_2{\rm F}_1\left(\gamma,\gamma;2\gamma;\frac{\bar q}{\bar k}\right) - (\gamma\to 1-\gamma).
\end{equation}
A factor antisymmetric under the replacement $\gamma \to 1-\gamma$
and as been absorbed in $\phi_0(\gamma,\vq).$ Note that in \eqref{eq:linvpr} we use the complex representation of vectors: $x\!=\!x_1\!+\! ix_2$ and $\bar x\!=\!x_1\!-\!ix_2$ for the vector $\x=(x_1,x_2);$ however from now on, we shall use $x$ to denote the modulus of the vector. Although this is not straightforward, one can explicitly check (see Appendix \ref{app:lin}) that \eqref{eq:linvpr} is an eigenfunction of the linear kernel of eq. \eqref{eq:fullbk} with the correct eigenvalue \eqref{eq:bfklchi}. When $k\gg q$, the hypergeometric functions go to 1 and $f^\gamma(\vk,\vq)$ becomes a simple power of $q/k$. The full solution of the linear part of \eqref{eq:fullbk} thus becomes
\[
\tN_{\text{lin}}(\vk,\vq) = \int_{\frac{1}{2}-i\infty}^{\frac{1}{2}+i\infty}
 \frac{d\gamma}{2i\pi} \phi_0(\gamma,\vq) \, \exp\left[\bar\alpha\chi(\gamma) Y - \gamma \log(k^2/q^2)\right],
\]
where irrelevant factors have again been absorbed in the initial condition $\phi_0(\gamma,\vq).$ 
This expression proves that the conditions for obtaining traveling waves and geometric scaling are fulfilled. More precisely, including all the terms common to each evolution equation of this universality class~\cite{Munier1}, it predicts
\begin{equation}\label{eq:newqs}
Q_s^2(Y) = q^2 \Omega_s^2(Y) \underset{Y\to\infty}{\sim} q^2
\exp\left[\bar\alpha\frac{\chi(\gamma_c)}{\gamma_c}Y-\frac{3}{2\gamma_c}\log(Y)
-\frac{3}{\gamma_c^2}\sqrt{\frac{2\pi}{\bar\alpha \chi''(\gamma_c)}} \frac{1}{\sqrt{Y}}
+{\cal{O}}(1/Y)
\right],
\end{equation}
together with the leading edge of the traveling front
\begin{equation}\label{eq:newfront}
\tN(\vk,\vq) \underset{Y\to\infty}{\sim}
\phi^{\gamma_c}(\vq)\,\log\left(\frac{k^2}{q^2\Omega_s^2(Y)}
\right)\,\abs{\frac{k^2}{q^2\Omega_s^2(Y)}}^{-\gamma_c}\,
\exp\left[-\frac{1}{2\bar\alpha\chi''(\gamma_c)Y}\log^2\left(
\frac{k^2}{q^2\Omega_s^2(Y)}\right)\right].
\end{equation}
When $q$ becomes smaller that the typical size $Q_T$ of the target, one can show that $Q_T$ substitutes to $q$ in the saturation scale. $Q_s$ is thus expected to be proportional to $q$ only at large $q$. The value of $\gamma_c$ is 0.6275. Note that those behaviours are the same as the ones predicted by the forward analysis. 

Let us also point out that all these results have been studied in terms of the momentum transfer $q$ and could not be derived in terms of the impact parameter $b$. One way to understand this is to look at the BFKL equation. One clearly sees that the equation in momentum space (see \eqref{eq:fullbk}) is local in $q$ while the equation in coordinate space (see \eqref{eq:bkspace}) couples different values of the impact parameter $b$. In the following section, we shall explicitly check that these predictions are observed by numerical simulations of the full equation \eqref{eq:fullbk}.

\section{Numerical analysis}\label{sec:numsol}

\subsection{Description of the method}

Instead of working directly with the momentum $\vk$, we shall use the 
variable $\vp=\vk-\vq/2$, which is the canonical conjugate of $\vrho$. 
The Fourier transform of the dipole amplitude 
then becomes
\[
\tN(\vp;\vq) = \frac{1}{(2\pi)^2} \int \frac{d^2r}{r^2}\,d^2b \,e^{i\vrho.\vp} 
\,e^{i\vecb.\vq} \N(\vrho;\vecb),
\]
and the BK equation in momentum space takes the form
\begin{eqnarray}\label{eq:fullbk_pq}
\partial_Y \tN(\vp;\vq) & = & 
\frac{\bar\alpha}{2\pi}\int\frac{d^2p'}{(\vp'-\vp)^2}
\left\{ 2\tN(\vp';\vq) 
  - \left[\frac{\left(\vp +\frac{\vq}{2}\right)^2}
               {\left(\vp'+\frac{\vq}{2}\right)^2 + (\vp'-\vp)^2}
        + \frac{\left(\vp -\frac{\vq}{2}\right)^2}
               {\left(\vp'-\frac{\vq}{2}\right)^2 + (\vp'-\vp)^2}
    \right] \tN(\vp;\vq)
\right\} \nonumber\\
 & - & \frac{\bar\alpha}{2\pi}\int d^2p'\, 
  \tN\left(\vp-\frac{\vp'}{2}+\frac{\vq}{4};\frac{\vq}{2}+\vp'\right)
  \tN\left(\vp-\frac{\vp'}{2}-\frac{\vq}{4};\frac{\vq}{2}-\vp'\right).
\end{eqnarray}
Although this formulation seems more complicated, it emphasises more clearly the 
symmetry properties of the BK equation under the transformation $\vk \to 
\vq-\vk$ which simply becomes $\vp \to - \vp$ in the new variables. 
Let us notice that the solution of the linear problem can be seen as depending 
only on the ratio $\vp/q$. Also, the limit $k \gg q$ we are interested 
in is equivalent to $p \gg q$.

For numerical studies, we shall therefore use $L=\log(Rp)$, $L_q = \log(Rq)$, where $R$ is the typical target size, and 
$\phi$, the angle between $\vp$ and $\vq$. It is obvious from 
\eqref{eq:fullbk_pq} that we have
\[
\tN(L, L_q, \phi) = \tN(L, L_q, -\phi) = \tN(L, L_q, \pi-\phi) = \tN(L, L_q, 
\pi+\phi),
\]
provided the same equality holds for the initial condition. In practice, we have computed the amplitude $\tN(L, L_q, \phi)$ on a lattice 
of $n_p \times n_q \times n_\phi$ points regularly spaced with 
$L_{\text{min}}\le L \le L_{\text{max}}$, $L_{q,\text{min}}\le L_q \le 
L_{q,\text{max}}$ and $0\le \phi \le \frac{\pi}{2}$.
We have then discretised the integral in the r.h.s. of \eqref{eq:fullbk_pq} and 
solved the resulting set of differential equations by a Runge-Kutta method.

\subsection{Results}

In this section, we shall study the properties of the numerical solutions of equation \eqref{eq:fullbk_pq}. We have studied two choices of initial distribution:
\begin{equation}\label{eq:init_exp}
\left.\tN(\vp;\vq)\right|_{Y=0} = \frac{1}{p^2+R^{-2}}\,e^{-Rq},
\end{equation}
and
\begin{equation}\label{eq:init_pert}
\left.\tN(\vp;\vq)\right|_{Y=0} = \frac{1}{p^2+R^{-2}}\frac{1}{1+q^2R^2}.
\end{equation}
Although the first choice seems more natural, the exponential damping at large $q$ leads to very small amplitudes producing numerical results difficult to handle. Therefore, in order to make the study at large $q$ easier, we shall also consider the less-steep behaviour \eqref{eq:init_pert}. The value of the coupling $\bar \alpha$ has been fixed to 0.2. We shall not consider the effect of the running of the strong coupling constant in this paper. 
Before presenting the results, let us notice that the dependence of $\tN$ on the angle $\phi$ is not really interesting for our purposes, hence, we shall concentrate on the amplitude averaged over $\phi$. Concerning the choice of the discretisation parameters, we have used 101 values of $L$ between $-5$ and 20, 53 values of $L_q$ between $-5$ and $8$ and 9 values of $\phi$ between 0 and $\pi/2$.

\begin{figure}
\subfigure[$\log(Rq)=-5$]{\includegraphics[scale=0.75]{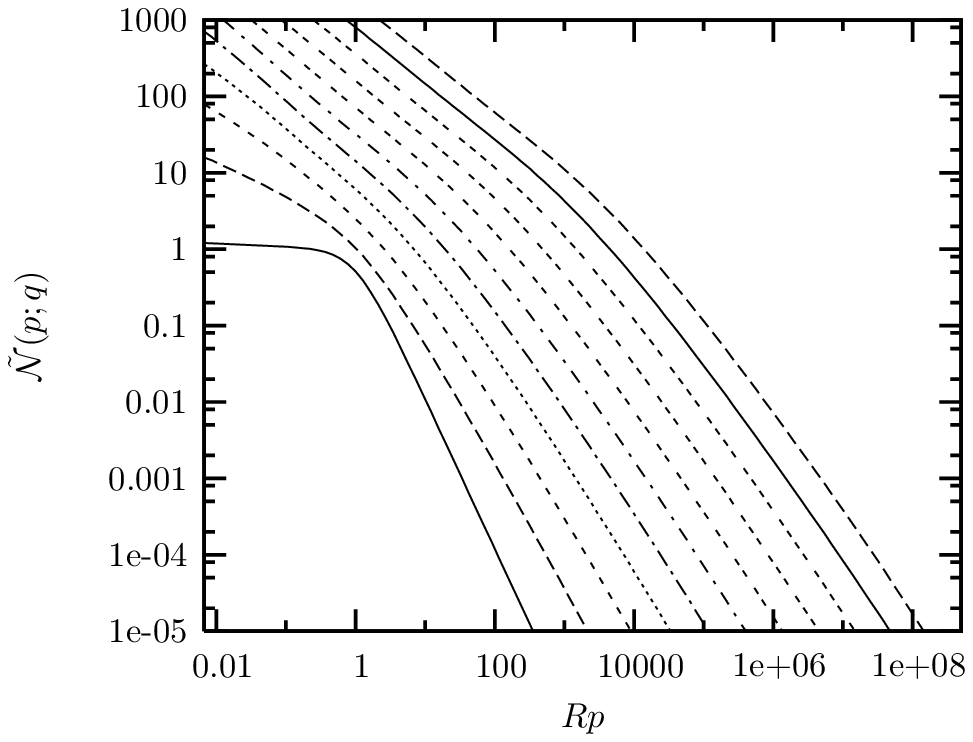}}\hspace{1.5cm}
\subfigure[$\log(Rq)=-1$]{\includegraphics[scale=0.75]{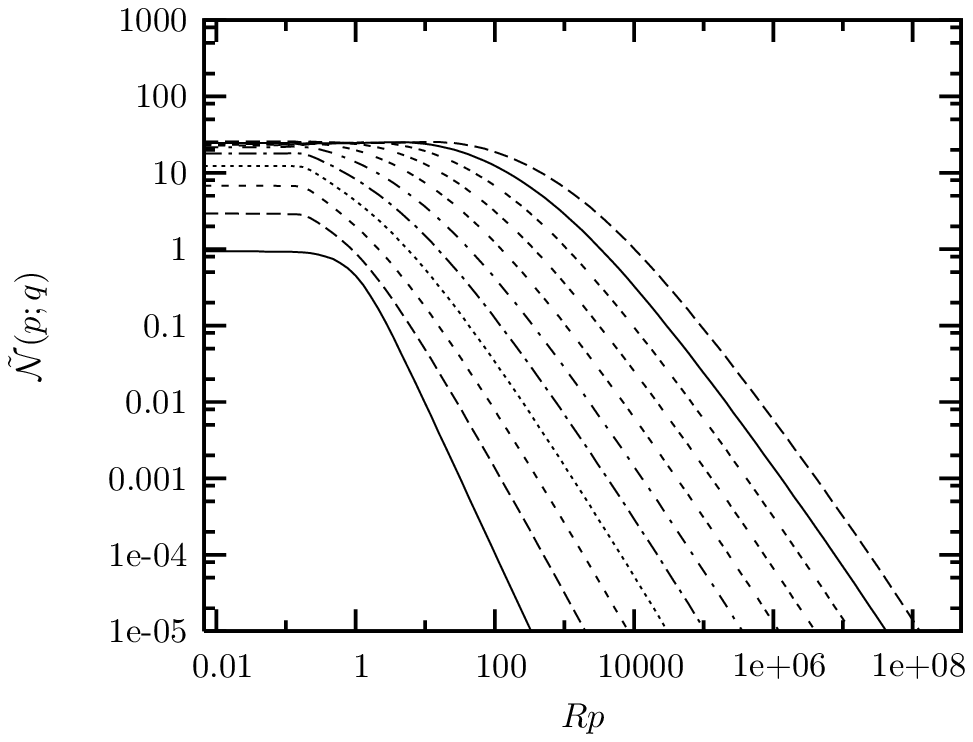}}
\subfigure[$\log(Rq)=0$]{\includegraphics[scale=0.75]{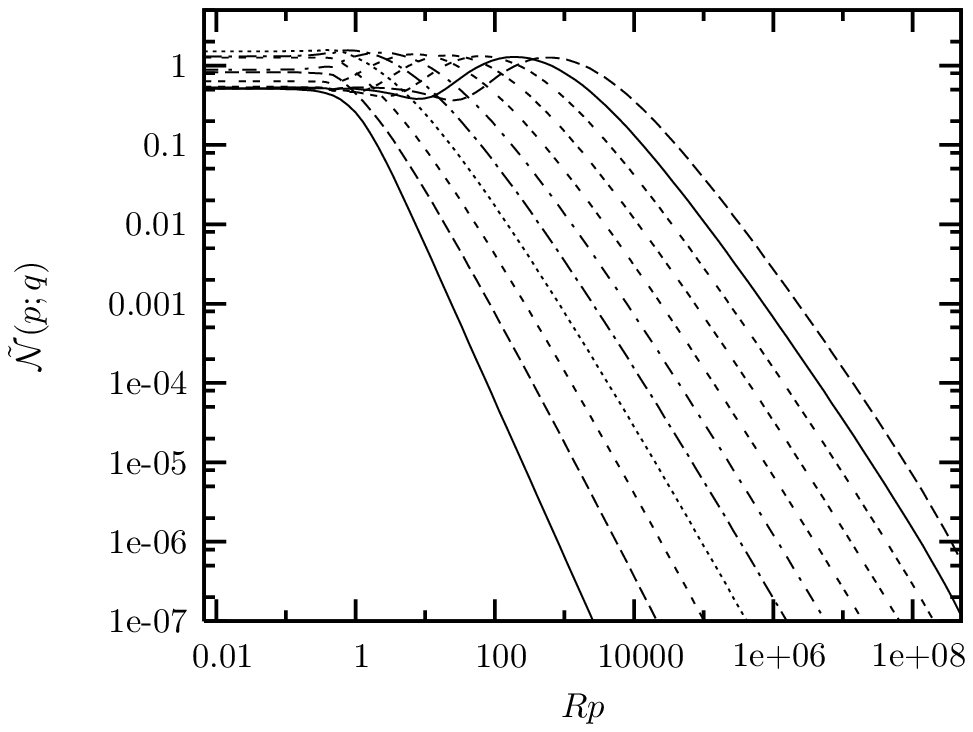}}\hspace{1.5cm}
\subfigure[$\log(Rq)=1$]{\includegraphics[scale=0.75]{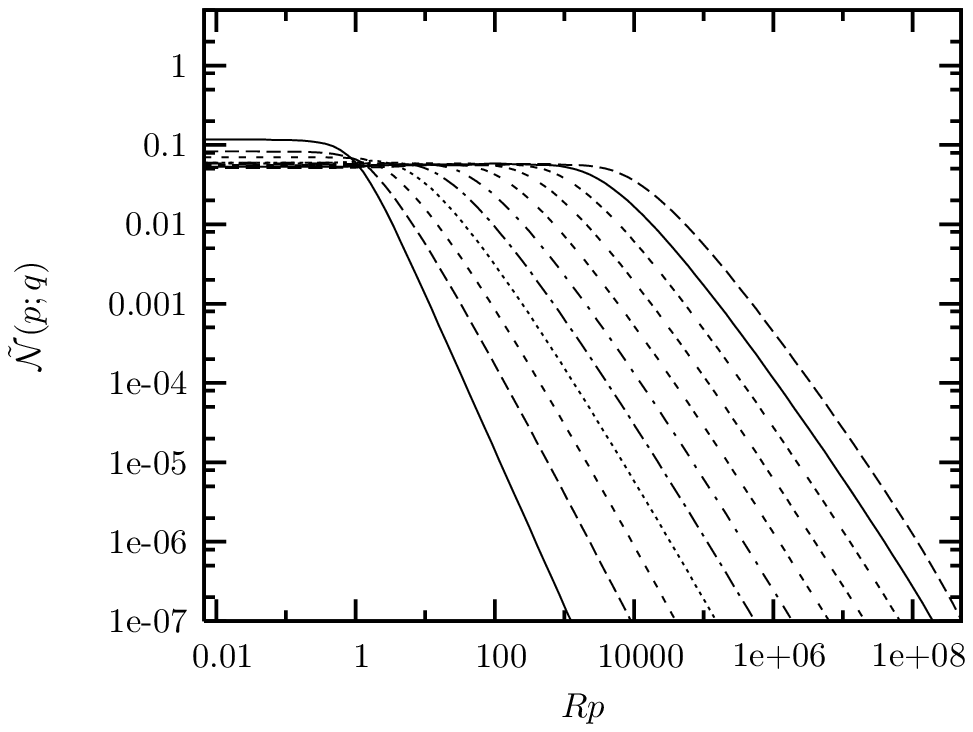}}
\subfigure[$\log(Rq)=2$]{\includegraphics[scale=0.75]{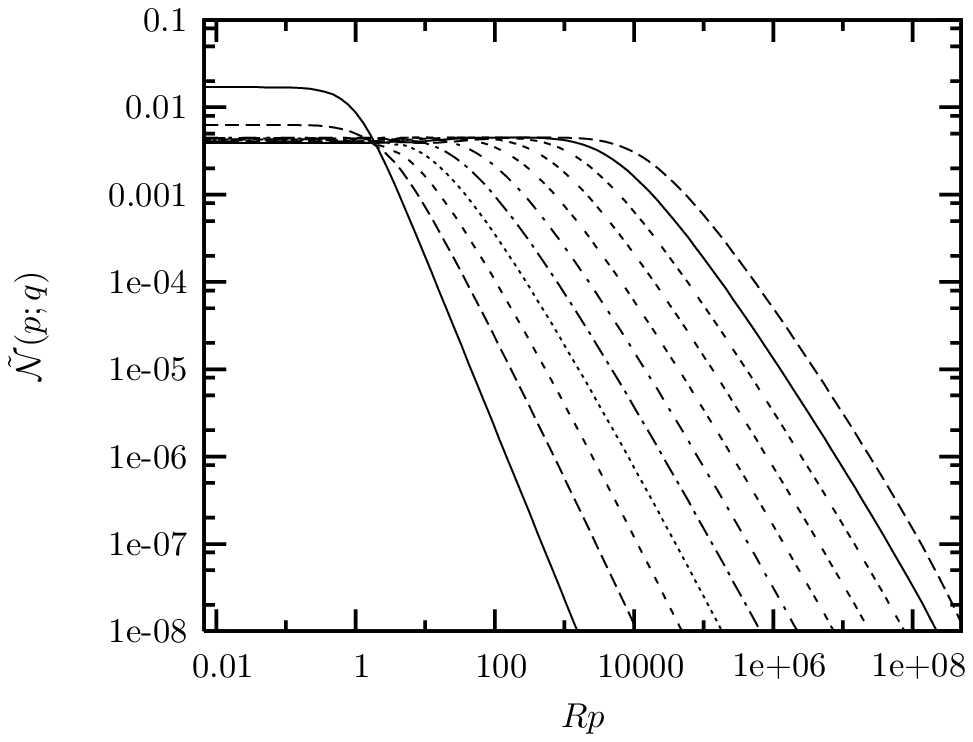}}\hspace{1.5cm}
\subfigure[$\log(Rq)=4$]{\includegraphics[scale=0.75]{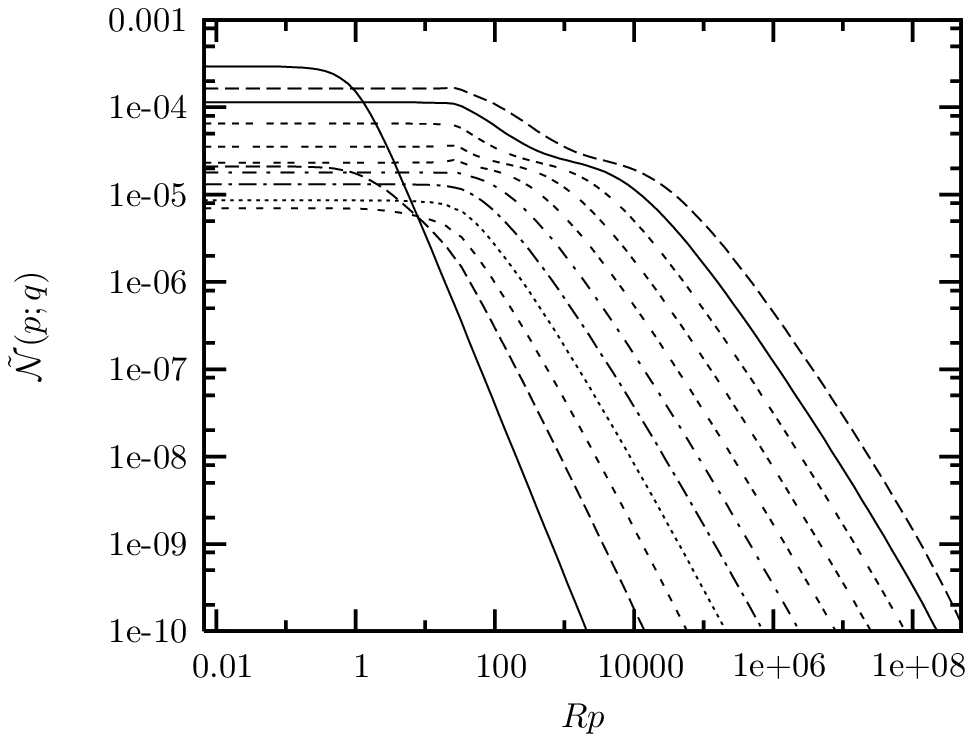}}
\caption{Rapidity evolution of the dipole density as a function of $p$ for different values of $q$. For each plot, we show the amplitude for $Y$ varying between 0 and 25 by steps of 2.5.}
\label{fig:qbins}
\end{figure}

To begin with, we shall check if the traveling waves predicted for $k \gg q$ in~\cite{us} are indeed observed in numerical simulations. This is done by plotting the rapidity evolution of the dipole amplitude as a function of $k$ (or, equivalently, $p$) for different values of $q$. As clearly seen in Figure \ref{fig:qbins}, as the rapidity increases, a traveling wave is formed at large values of $p$. For these results, the initial condition used is \eqref{eq:init_pert}. 

\begin{figure}
\subfigure[]{\includegraphics[scale=0.9]{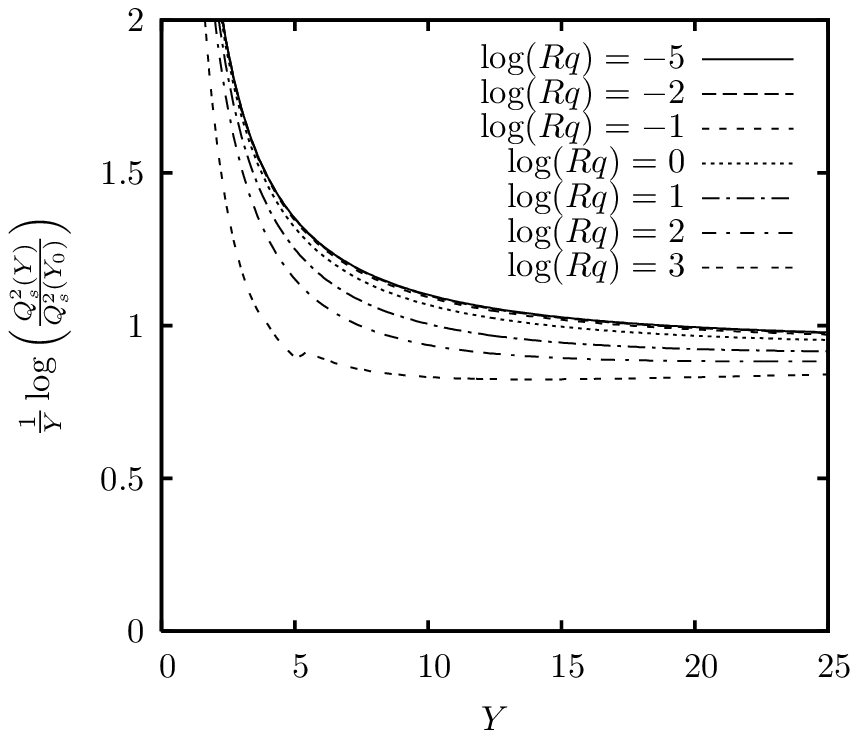}}
\subfigure[]{\includegraphics[scale=0.9]{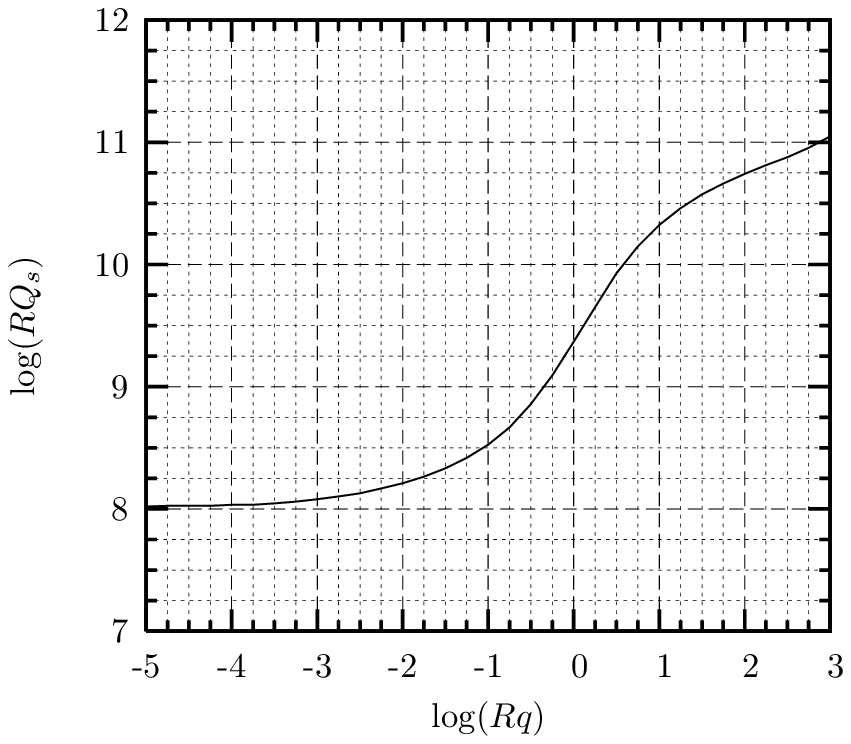}}
\caption{This figure presents the results concerning the saturation scale: (a) shows that the evolution of the saturation scale with $Y$ has the predicted behaviour and (b) represents the $q$-evolution of the saturation scale (see text for details).}
\label{fig:qs}
\end{figure}

To observe this more precisely, it is interesting to study the properties of the saturation scale $Q_s$. In~\cite{us}, it was predicted that $Q_s$ should evolve asymptotically like a $q$-independent power of the energy, and, at fixed rapidity, should grow linearly with $q$ when $q$ becomes larger than the inverse size of the target. Let us first check the high-energy dependence of the saturation scale. Due to the $q$-dependence of the initial condition \eqref{eq:init_pert}, one cannot simply extract the saturation scale by solving $\tN(p=Q_s(Y,q);q)=\bar N_0$ since, for sufficiently large values of $q$, the amplitude will always be smaller than $\bar N_0$. To fix this problem, we rely on the fact that the asymptotic behaviour of $\log(Q_s)$ is expected to be insensitive to the choice of $\bar N_0$. We shall thus use a $q$-dependent threshold $\bar N_0(q)$, taking into account the damping at large $q$ introduced by the initial condition. This leads us to solve
\begin{equation}\label{eq:qsdef}
\tN(p=Q_s(Y,q);q)=\bar N_0(q)=\frac{\bar N_0(0)}{1+R^2 q^2},
\end{equation}
for different values of $Y$ and $q$. In Figure \ref{fig:qs}(a), where we have chosen $\bar N_0(0)=0.05$, we show that whatever $q$ is, $\log(Q_s(Y)/Q_s(Y_0))$, with $Y_0$ fixed to 1, goes to a constant value at asymptotic rapidities. In fact, one can even check that our numerical results are consistent with the expansion
\[
\partial_Y \log\left[Q_s^2(Y,q)\right] = \bar\alpha\frac{\chi(\gamma_c)}{\gamma_c} - \frac{3}{2\gamma_c}\frac{1}{Y}+\frac{3}{\gamma_c^2}\sqrt{\frac{2\pi}{\bar\alpha\chi''(\gamma_c)}}\frac{1}{Y^{3/2}}+\frac{c_2(q)}{Y^2},
\]
valid at large $Y$. Notice that, in this expression, the $q$-dependent non-universal term $c_2 Y^{-2}$ is required in the rapidity range under consideration.

The extraction of the $q$-dependence of the saturation scale is more tricky since the definition \eqref{eq:qsdef} is obviously not sufficient. Basically, one has to obtain the position of the wavefront in a way that does not depend on any threshold $\bar N_0$ at all. Our method consists in studying the so-called {\em reduced front} $p^{2\gamma_c} \tN(p,q)$ which is expected to show a maximum in $p$ in the region where geometric scaling holds~\cite{Munier1}. However, since the violations of geometric scaling in the ultra-violet appear outside of our lattice, we shall define $Q_s$ as the maximum of a {\em modified version} of the reduced front $p \tN(p,q)$ {\em i.e.} replace $\gamma_c=0.6275$ by 0.5. Physically, the maximum of this quantity can be considered as a measure of the momentum for which the saturated regime is replaced by the traveling front. We have applied this method to extract the average\footnote{We average over an interval of rapidity in order to minimise numerical uncertainties.} saturation scale for $20\le Y \le 25$ and checked that the result does not change if one replace 0.5 by another value close to it. The result is shown in Figure \ref{fig:qs}(b) and we see that the saturation scale obtained from this method is constant for small momentum transfer and shows a linear increase as $q$ becomes larger than $R^{-1}$. For very high values of $q$ ($\log(Rq) \gtrsim 2$), the results start to diverge from this proportionality. We attribute this feature to the fact that the asymptotic traveling-wave behaviour is not yet reached, and at those high values of $q$, we are not extracting the true saturation scale.

\begin{figure}
\includegraphics{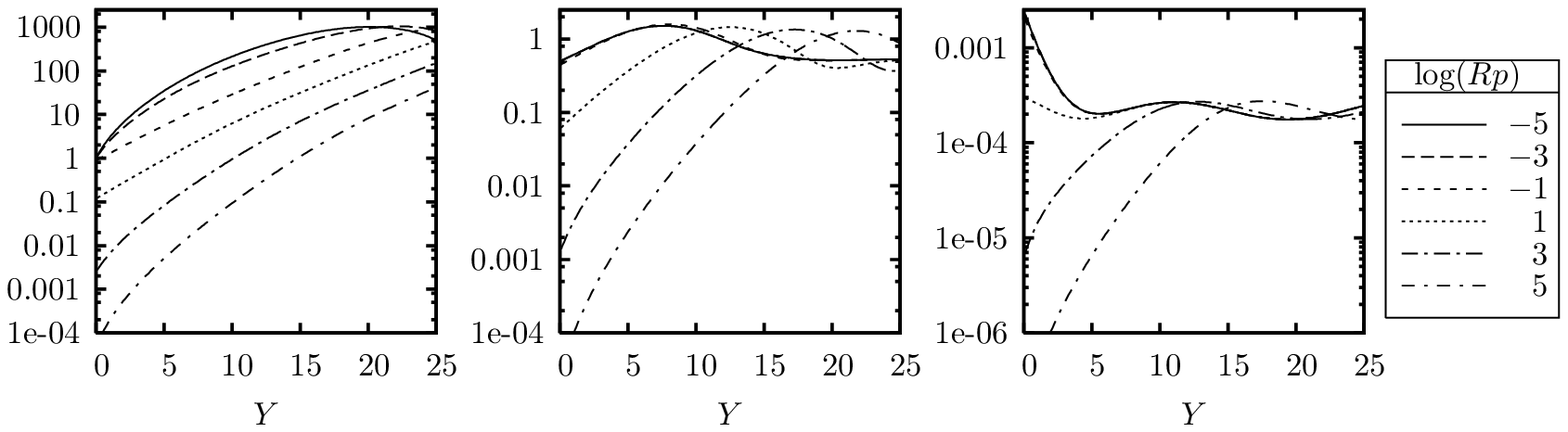}
\caption{This figure shows the $Y$ dependence of the amplitude $\tN(\vp;\vq)$. In the left plot we have fixed $\log(Rq)=-3$, while $\log(Rq)=0$ in the centre plot and $\log(Rq)=3$ in the right one.}
\label{fig:ydep}
\end{figure}

\begin{figure}
\includegraphics{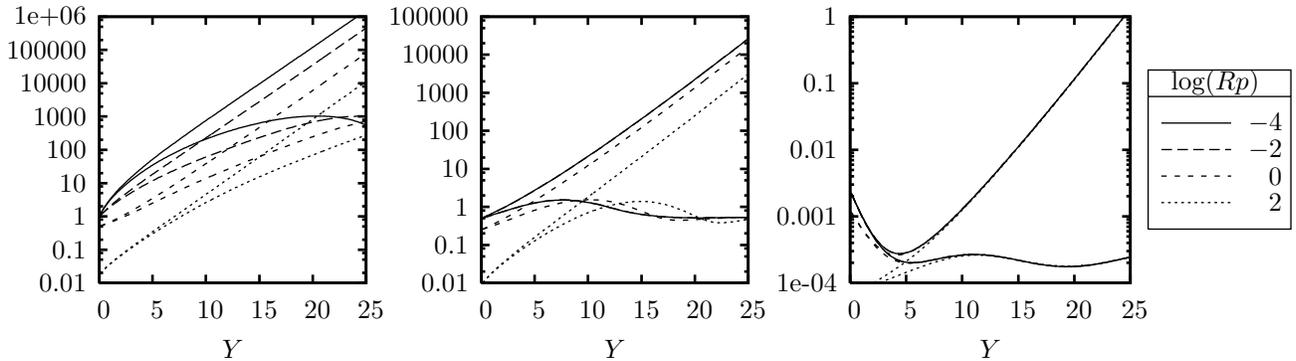}
\caption{This figure shows the $Y$ dependence of the amplitude $\tN(\vp;\vq)$ compared to the amplitude obtained by BFKL evolution. In each pair of curves, the upper one is obtained from BFKL evolution and the lower one includes nonlinear effects. In the left plot we have fixed $\log(Rq)=-3$, while $\log(Rq)=0$ in the centre plot and $\log(Rq)=3$ in the right one.}
\label{fig:vsbfkl}
\end{figure}

\begin{figure}
\includegraphics{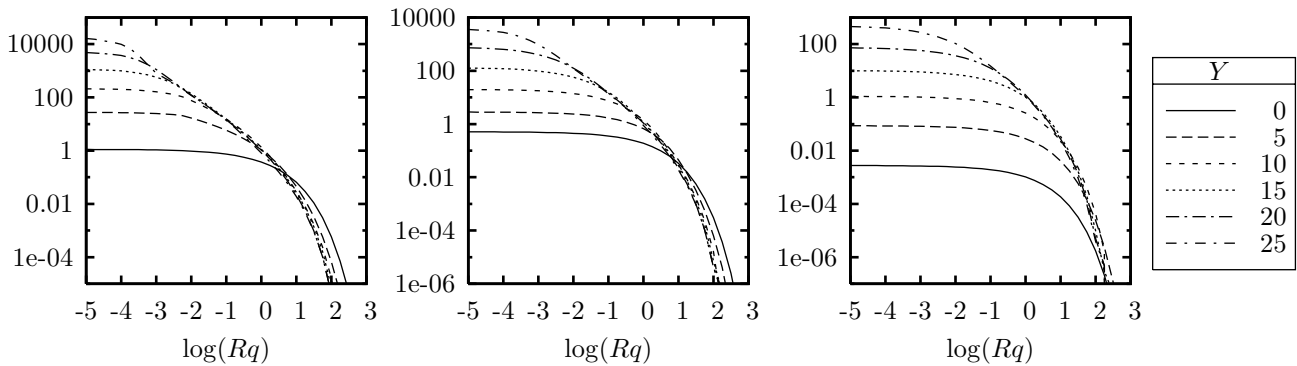}
\caption{Rapidity evolution of the dipole density as a function of $q$ for different values of $p$. In the left (resp. centre, right) plot we have set $\log(Rp)=-3$ (resp. 0, 3).}
\label{fig:kbins}
\end{figure}

Let us say a few words about the evolution in the regime $k\!\sim\!q$. As shown in Figure \ref{fig:ydep}, this region presents oscillations and the amplitude seems to saturate to a constant value at large $Y$. To see more precisely the effect of the nonlinear term in \eqref{eq:fullbk}, we can compare the $Y$-evolution of the amplitude presented in figure \ref{fig:ydep}, with the results obtained from BFKL evolution {\em i.e.} keeping only the linear term in \eqref{eq:fullbk}. The comparison of these two evolutions is presented in Figure \ref{fig:vsbfkl}. This figure shows interesting features. First, one clearly sees the damping due to the inclusion of the nonlinear effects. Then, one observes the exponential growth of the BFKL amplitude with rapidity. Finally, we see that at high energy and for a given value of $q$, the BFKL amplitude becomes degenerate for $k\ll q$. This feature, closely related to the plateau observed in the infrared region in Figure \ref{fig:qbins}, shows that $q$ acts like an infrared cut-off \cite{mueltri}.

Finally, let us see how the $q$ behaviour of the amplitude evolves with rapidity. In Figure \ref{fig:kbins}, we show how the initial condition \eqref{eq:init_exp} evolves with rapidity for different values of $p$, as a function of $q$. First of all, it appears that the extension in $q$ remains roughly constant as the rapidity increases. Then, we see an increase of the amplitude for small values of $q$. As noticed in~\cite{us}, the fact that the $q$-behaviour of the amplitude does not develop a power-law tail when the energy increases as it is the case with impact parameter, is one of the main advantage in studying the BK equation in momentum space.


\section{Conclusion}\label{sec:ccl}

Let us summarize the main results of this paper. First, we have proposed a Fourier transform of the dipole scattering amplitude which leads to a new form of the BK high-energy evolution equation. This new form involves the amplitude in momentum space instead of coordinate space (see equation \eqref{eq:fullbk}). We have shown that the momentum-transfer-dependent equation reduces to the usual momentum-space evolution in the $b$-independent case. This equation is also obtained in the forward case $q=0$, assuming a steep $q$ dependence of the amplitude. 

We have then checked numerically the properties of the asymptotic solutions of the proposed $q$-dependent BK equation. We have shown that all the properties predicted theoretically in~\cite{us} are indeed reproduced by numerical simulations. More precisely, one obtains traveling waves at large rapidities, leading to geometric scaling in the region $k \gg q$. The speed of the traveling front converges to the $q$-independent value predicted by previous analysis. Concerning the saturation scale, we have shown that it goes from a constant behaviour to a linear increase with momentum transfer when $q$ goes from values smaller than the target typical scale to larger values. While the coordinate-space version of the BK equation leads to a diffusion in impact parameter already at the level of the linear kernel, the corresponding equation in momentum space remains local in $q.$ This means that, to the contrary of what happens in impact-parameter space where the evolution generates a perturbative tail at large $b$, the $q$ behaviour does not show an increase at large values of $q$. This important property is also the reason for which momentum space is the natural frame to discuss traveling-wave solutions and geometric scaling.


In addition to these results, the possibility to solve the full BK equation allows for various extensions. From the phenomenological point of view, it can help to test saturation effects in many processes including {\em e.g.} diffractive photoproduction of vector mesons. On more theoretical grounds, one may be tempted to analyse the odderon equations Ref.\cite{Hatta:2005as}. Due to the parity of the odderon, this contribution cannot be introduced at the level of the $b$-independent BK equation. In all these cases, it would be interesting to introduce the effect of the running coupling. Finally, this equation can also serve as a basis for studies of fluctuation effects and the corresponding Langevin equation derived very recently~\cite{stoc}.

\begin{acknowledgments}
We would like to thank R. Peschanski for fruitful discussions and suggestions. G.S. is funded by the National Funds for Scientific Research (Belgium).
\end{acknowledgments}

\begin{appendix}

\section{Solutions of the linearised equation}\label{app:lin}

In this appendix, we write down the main lines that show that \eqref{eq:linvpr} is a solution of the linear part of equation \eqref{eq:fullbk} with the correct eigenvalue. We start by going to the complex representation and use the integral representation of $f^\gamma(\vk,\vq),$ see Appendix B of~\cite{us}. One has, up to a $\gamma$-dependent factor:
\[
f^\gamma(x) = \abs{x}^{2\gamma}\int d^2v\,\abs{v}^{2\gamma-2} \abs{1-v}^{2\gamma-2} \abs{1-\bar x v}^{-2\gamma},
\]
with $x=q/k$. If we perform the change of variables $k' \to v=k'/k$ in (the linear part of) \eqref{eq:fullbk}, we are left with the integral
\[
I^\gamma(x) = \frac{\bar\alpha}{2\pi} \int\frac{d^2v}{\abs{1-v}^2}\left[ 2f^\gamma(x/v)-\frac{1}{\abs{v}^2}f^\gamma(x)\right] + \frac{\bar\alpha}{4\pi} \int\frac{d^2v}{\abs{1-v}^2}\left(\frac{1}{\abs{v}^2}-\frac{\abs{1-x}^2}{\abs{v-x}^2}\right) f^\gamma(x).
\]
The second integral can be computed, {\em e.g.} by dimensional regularisation. We also perform the change of variables $v\to u=1/v$ in the first integral, leading to
\[
I^\gamma(x) = \frac{\bar\alpha}{\pi} \int\frac{d^2u}{\abs{1-u}^2}\left[ \frac{1}{\abs{u}^2} f^\gamma(ux)-\frac{1}{1+\abs{1-u}^2}f^\gamma(x)\right] + \frac{\bar\alpha}{2} \log\left(\frac{1}{\abs{1-x}^2}\right) f^\gamma(x).
\]
Inserting the integral representation of $f^\gamma(x)$, one obtains
\begin{eqnarray*}
I^\gamma(x) & = & \frac{\bar\alpha}{\pi} \abs{x}^{2\gamma}
\int d^2v\,\abs{v}^{2\gamma-2} \abs{1-v}^{2\gamma-2} 
\lim_{\varepsilon\to 0} \left(
\int\frac{d^2u}{\abs{1-u}^{2-2\varepsilon}} \abs{u}^{2\gamma-2}\abs{1-\bar vxu}^{-2\gamma}
-\int\frac{d^2u}{\abs{1-u}^{2-2\varepsilon}}\frac{\abs{1-\bar v x}^{-2\gamma}}{1+\abs{1-u}^2} 
\right)\\
& + & \frac{\bar\alpha}{2} \log\left(\frac{1}{\abs{1-x}^2}\right) f^\gamma(x).
\end{eqnarray*}
where we have explicitly regularised the divergence at $u=1$.

Performing the integration over $u$ and taking the limit $\varepsilon\to 0$, we get
\[
I^\gamma(x) = \bar\alpha\chi(\gamma)f^\gamma(x) + \bar\alpha R^\gamma(x),
\]
with
\begin{eqnarray*}
R^\gamma(x) & = & \frac{1}{2} \log\left(\frac{1}{\abs{1-x}^2}\right) f^\gamma(x) \\
& + & \abs{x}^{2\gamma}\int d^2v\,\abs{v}^{2\gamma-2}\abs{1-v}^{2\gamma-2}
\left[
(1-\bar x v)^{-\gamma} \partial_\varepsilon\, {}_2F_1(\gamma,\gamma;\gamma+\varepsilon;\bar v x)
+ 
(1-x \bar v)^{-\gamma} \partial_\varepsilon\, {}_2F_1(\gamma,\gamma;\gamma+\varepsilon;v \bar x)
\right]_{\varepsilon=0}.
\end{eqnarray*}
The derivative of the hypergeometric function can be explicitly computed and we are finally left with the integration
\[
R^\gamma(x) = \frac{\abs{x}^{2\gamma}}{2}
\int d^2v\,\abs{v}^{2\gamma-2}\abs{1-v}^{2\gamma-2}\abs{1-\bar xv}^{-2\gamma}
 \log\left(\frac{\abs{1-\bar xv}^4}{\abs{1-x}^2}\right).
\]
Finally, the conformal transformation $v=\frac{1-u}{1-\bar x u}$ gives
\[
R^\gamma(x) = \frac{\abs{x}^{2\gamma}}{2}
\int d^2u\,\abs{u}^{2\gamma-2}\abs{1-u}^{2\gamma-2}\abs{1-\bar xu}^{-2\gamma}
 \log\left(\frac{\abs{1-x}^2}{\abs{1-\bar xu}^4}\right)= -R^\gamma(x).
\]
Therefore, $R^\gamma(x)=0$ and $f^\gamma(x)$ is an eigenfunction of the linear kernel with $\chi(\gamma)$ as associated eigenvalue.

\end{appendix}


\begin{thebibliography}{99}

\bibitem{bfkl}
L. N. Lipatov, {\it Sov. J. Nucl. Phys.} {\bf 23}, (1976) 338;
E. A. Kuraev, L. N. Lipatov and V. S. Fadin,
{\it Sov. Phys. JETP} {\bf 45}, (1977) 199;
I. I. Balitsky and L. N. Lipatov,
{\it Sov. J. Nucl. Phys.} {\bf 28}, (1978) 822.

\bibitem{glr} L. V. Gribov, E. M. Levin and M. G. Ryskin, {\it Phys. Rep.}
{\bf 100} (1983) 1.

\bibitem{glr+} 
A. H. Mueller and J. Qiu, {\it Nucl. Phys.} {\bf B268} (1986) 427;
E. Levin and J. Bartels, {\it Nucl. Phys.} {\bf B387} (1992) 617.

\bibitem{dipole}
N. N. Nikolaev and B. G. Zakharov, {\it Zeit. f\"ur. Phys.} {\bf C49} (1991) 
607; {\it Phys. Lett.} {\bf B332} (1994) 184.

\bibitem{mueller}
A. H. Mueller, {\it Nucl. Phys.} {\bf B415} (1994) 373;
A. H. Mueller and B. Patel, {\it Nucl. Phys.} {\bf B425} (1994) 471;
A. H. Mueller, {\it Nucl. Phys.} {\bf B437} (1995) 107.

\bibitem{kov}
Y. V. Kovchegov, {\it Phys. Rev.} {\bf D60} (1999) 034008;
{\it Phys. Rev.} {\bf D61} (2000) 074018.

\bibitem{bal}
I. Balitsky, {\it Nucl. Phys.} {\bf B463} (1996) 99.


\bibitem{bk+} 
E. Levin and K. Tuchin, {\it Nucl. Phys.} {\bf A691} (2001) 779;
{\it Nucl. Phys.} {\bf A693} (2001) 787.

\bibitem{Munier1}
S.~Munier and R.~Peschanski,
Phys.\ Rev.\ Lett.\  {\bf 91}, 232001 (2003)
[arXiv:hep-ph/0309177]; 
Phys.\ Rev.\ {\bf D69}, 034008 (2004)
[arXiv:hep-ph/0310357]; 
Phys.\ Rev.\ {\bf D70}, 077503 (2004)
[arXiv:hep-ph/0310357]. 

\bibitem{geomsc} 
A.~M.~Stasto, K.~Golec-Biernat and J.~Kwiecinski,
Phys.\ Rev.\ Lett.\  {\bf 86} (2001) 596 [arXiv:hep-ph/0007192].

\bibitem{conf}
A.~Kovner and U.~A.~Wiedemann,
Phys.\ Rev.\ D {\bf 66} (2002) 051502 [arXiv:hep-ph/0112140],
Phys.\ Lett.\ B {\bf 551} (2003) 311 [arXiv:hep-ph/0207335];
E.~Ferreiro, E.~Iancu, K.~Itakura and L.~McLerran,
Nucl.\ Phys.\ A {\bf 710} (2002) 373 [arXiv:hep-ph/0206241].

\bibitem{bknum}
K.~Golec-Biernat and A.~M.~Stasto,
Nucl.\ Phys.\ B {\bf 668} (2003) 345 [arXiv:hep-ph/0306279],
E.~Gotsman, M.~Kozlov, E.~Levin, U.~Maor and E.~Naftali,
Nucl.\ Phys.\ A {\bf 742} (2004) 55 [arXiv:hep-ph/0401021].

\bibitem{us}
C.~Marquet, R.~Peschanski and G.~Soyez,
arXiv:hep-ph/0502020.

\bibitem{bknumimp}
M.~Braun,
Eur.\ Phys.\ J.\ C {\bf 16} (2000) 337 [arXiv:hep-ph/0001268],
Phys.\ Lett.\ B {\bf 576} (2003) 115 [arXiv:hep-ph/0308320];
K.~Golec-Biernat, L.~Motyka and A.~M.~Stasto,
Phys.\ Rev.\ D {\bf 65} (2002) 074037 [arXiv:hep-ph/0110325];
J.~L.~Albacete, N.~Armesto, J.~G.~Milhano, C.~A.~Salgado and U.~A.~Wiedemann,
Phys.\ Rev.\ D {\bf 71} (2005) 014003 [arXiv:hep-ph/0408216].

\bibitem{KPP}
R.~A. Fisher,
\newblock Ann. Eugenics {\bf 7}, 355 (1937);
\newblock A.~Kolmogorov, I.~Petrovsky, and N.~Piscounov,
\newblock Moscou Univ. Bull. Math. {\bf A1}, 1 (1937).

\bibitem{Lipatov86} 
L.~N.~Lipatov,
Sov.\ Phys.\ JETP {\bf 63}, 904 (1986)
[Zh.\ Eksp.\ Teor.\ Fiz.\  {\bf 90}, 1536 (1986)].

\bibitem{NP} H.~Navelet and R.~Peschanski,
Nucl.\ Phys.\ {\bf B507}, 353 (1997)
[arXiv:hep-ph/9703238].

\bibitem{Navelet:1997tx}
H.~Navelet and S.~Wallon,
Nucl.\ Phys.\ {\bf B522}, 237 (1998)
[arXiv:hep-ph/9705296].

\bibitem{LipatovReview}
L.~N.~Lipatov,
Phys.\ Rept.\ {\bf 286}, 131 (1997)
[arXiv:hep-ph/9610276].

\bibitem{mueltri}
A.~H.~Mueller and D.~N.~Triantafyllopoulos,
Nucl.\ Phys.\ B {\bf 640} (2002) 331
[arXiv:hep-ph/0205167].

\bibitem{Hatta:2005as}
Y.~V.~Kovchegov, L.~Szymanowski and S.~Wallon,
Phys.\ Lett.\ B {\bf 586} (2004) 267
[arXiv:hep-ph/0309281];
Y.~Hatta, E.~Iancu, K.~Itakura and L.~McLerran,
arXiv:hep-ph/0501171.

\bibitem{stoc}
A.~H.~Mueller and A.~I.~Shoshi, Nucl.\ Phys.\ B {\bf 692} (2004) 175
[arXiv:hep-ph/0402193];
E.~Iancu, A.~H.~Mueller and S.~Munier, Phys.\ Lett.\ B {\bf 606} (2005) 342
[arXiv:hep-ph/0410018];
E.~Iancu and D.~N.~Triantafyllopoulos, arXiv:hep-ph/0411405,
Phys.\ Lett.\ B {\bf 610} (2005) 253 [arXiv:hep-ph/0501193];
A.~H.~Mueller, A.~I.~Shoshi and S.~M.~H.~Wong, arXiv:hep-ph/0501088.

\end{thebibliography}
\end{document}